\documentclass[twocolumn,superscriptaddress,amsmath,amssymb,aps,pra]{revtex4}
\usepackage{graphicx}
\usepackage{amsmath}
\usepackage{amssymb}
\usepackage{braket}
\usepackage{dcolumn}
\usepackage[colorlinks=true,linkcolor=blue,citecolor=blue,urlcolor=blue]{hyperref}

\begin{document}

\title{Interaction Control of Ultracold Alkaline-Earth Atoms}
\author{Ren Zhang}
\affiliation{School of Science, Xi'an Jiaotong University, Xi'an, 710049, China}
\author{Yanting Cheng}
\affiliation{Institute for Advanced Study, Tsinghua University, Beijing, 100084, China}
\author{Peng Zhang}
\affiliation{Department of Physics, Renmin University of China, Beijing, 100872,
China}
\author{Hui Zhai}
\email{hzhai@tsinghua.edu.cn}
\affiliation{Institute for Advanced Study, Tsinghua University, Beijing, 100084, China}

\date{\today}

\begin{abstract}
Ultracold alkaline-earth atoms have now been widely explored for precision measurements and quantum simulation. Because of its unique atomic structure, alkaline earth atoms possess great advantages for quantum simulation and studying quantum many-body matters, such as simulating synthetic gauge field, Kondo physics and $SU(N)$ physics. To fully explore the potential of ultracold alkaline-earth atoms, these systems also need to be equipped with the capability of tuning the inter-atomic interaction to the strongly interacting regime. Recently several theoretical proposals and experimental demonstrations have shown that both spin-independent and spin-exchanging interaction can be tuned to resonance. In this perspective, we will review these progress and discuss the new opportunities brought by these interaction control tools for future quantum simulation studies with ultracold alkaline-earth atoms.   

\end{abstract}

\maketitle

\section{Introduction \label{Introduction}}

In the past two decades, quantum simulation with cold atom physics has greatly enriched our understanding of macroscopic quantum matters \cite{RMP_MB,NP_QG,OL_review,Zoller_review}. The rapid progresses of cold atom physics benefit from new ideas and the development of new technologies. Moreover, it benefits from the fact that new elements of atoms are cooled to quantum degeneracy and join the ultracold family. At the beginning of this field, only alkaline-metal atoms are explored in experiments. Later experimentalists are able to cool alkaline-earth (AE) atoms \cite{Ybnote} into quantum degeneracy \cite{Ybbec,Yb173,Cabec,Sr84_1,Sr84_2,Sr87} and these atoms are now used more and more often in many laboratories. It has now been realized that ultracold AE atoms have unique advantages for quantum simulation and for studying quantum many-body physics. To start with, we first highlight two important features of AE atoms due to the special electronic structure of AE atoms.  

\begin{itemize}
  \item {\bf Two-Orbital Physics:} The outmost $s$-orbital of an AE atom is occupied by two electrons, thus for the electronic ground state labeled by ${}^1S_0$, both the total electron spin and the total angular momentum are zero. For the electronically excited state labeled by ${}^3P_0$, one of the two electrons is excited to $p$-orbital and the total electron spin $S=1$, and because its electronic spin is different from the ground state electronic spin, the dipolar transition between ${}^3P_0$ and ${}^1S_0$ is forbidden. As a result, the spontaneous emission is strongly suppressed. For this reason, the single-atom lifetime of ${}^3P_0$ state can be as long as seconds \cite{lifetime}. When we consider a mixture of atoms in both ${}^1S_0$ and ${}^3P_0$ states, we introduce an orbital degree of freedom to distinguish them, and these two states are now referred to as a doublet of the orbital degree of freedom.
  
This two-orbital physics is very useful. For precision measurement, the optical transition between these two states is now used for building the best optical clock \cite{clock1,clock_review,clock2,clock3,clock4,clock5,clock6}, with which the current world record for the clock accuracy of $10^{-19}$ has been achieved \cite{clock6}. Recently, the optical coupling between two orbital states has also been used for quantum simulation of spin-orbit coupling effect \cite{socclock1,socclock2,socfootnote}, and the advantage is that it can largely avoid the heating problem due to the spontaneous emission. The heating caused by spontaneous emission has been the main obstacle preventing realizing many interesting many-body physics when the spin-orbit coupling is simulated in degenerate alkali-metal atomic gases \cite{socreviewsp,socreviewhui}. 
    
  \item {\bf $SU(N)$ Spin Physics:} The fermionic AE atoms also have non-zero nuclear spin $I$, and in particular, the nuclear spin is quite large for many isotopes. For ${}^1S_0$ and ${}^3P_0$ states, since the electron angular momentum $J=0$, the absence of hyperfine coupling leads to that the interactions between atoms are independent of the nuclear spins \cite{XiboSUN,leoSUN,simonSUN}. Therefore, at zero magnetic field, the system possesses a $SU(N)$ symmetry with $N=2I+1$.  

\end{itemize}    

These two features discussed above are unique to AE atoms which can unlock many new opportunities for quantum simulation. Moreover, on the technical side, since there is a much longer history of using AE atoms for building an optical clock, and many technologies that have been developed there can be naturally transferred to the quantum simulation study for performing very precise measurement of many-body physics with AE atoms, for instance, to measure the interaction energy with very high precision \cite{measure-int1,measure-int2,measure-int3,measure-int4,measure-int5,measure-int6}. Because of these reasons, ultracold AE atoms have now become a major platform for quantum simulation studies. 

Another important aspect of cold atom physics is the ability to control interactions through Feshbach resonance (FR) \cite{FR1,FR2} (See Box). The progress made with alkaline-metal atoms in the previous two decades is largely benefited from this tool of interaction control \cite{Fermigas_review,Strinati}. In particular, when the interaction energy can be tuned to be comparable or even larger than the kinetic energy, the conventional perturbation theory fails to capture the strong correlation between particles, where the physics is not yet well understood, and hence the approach of quantum simulation becomes very valuable. Therefore, to fully explore the potential of AE atoms in quantum simulation, it is also highly desirable to develop tools to control interactions between AE atoms. In alkali-metal atoms, two of the most widely used methods for tuning the interactions are the magnetic FR \cite{FR1,FR2} and the confinement-induced resonance(CIR) \cite{oshanii98}. In this perspective, we will focus on two related recent developments in AE atoms. 

First, because the electronic spin $J=0$ for both ${}^1S_0$ and ${}^3P_0$ states and the magnetic FR relies on the electronic spin, it does not exist in ${}^1S_0$ and ${}^3P_0$ states of AE atoms. Nevertheless, we will describe an alternative proposal to use the ``orbital degree of freedom" to replace the role of the electronic spin, with which the nuclear-spin-independent interaction can be tuned by the magnetic field strength \cite{OFR}. This proposal has been realized by the Munich group \cite{OFR_exp1} and the Florence group \cite{OFR_exp2}, and now it has been used to explore strongly interacting physics in ultracold AE atoms \cite{polaron-exp}. Secondly, the CIR also works for AE atoms, and it turns out that we can utilize the CIR to tune the nuclear-spin exchanging interaction, which can enhance the Kondo effect \cite{CIR1,CIR2}. This effect has also been observed recently by the Munich group \cite{CIR_exp}. In this perspective, we will first review the developments of controlling both the spin-independent and spin-dependent-exchanging interactions of AE atoms, respectively, and then we will discuss future opportunities for quantum simulation and quantum many-body physics brought by combining these interaction control tools with two-orbital and $SU(N)$ physics.  

\section{Interaction between Alkaline-Earth Atoms}

\begin{figure*}
\begin{center}
\includegraphics[width=0.5\textwidth]{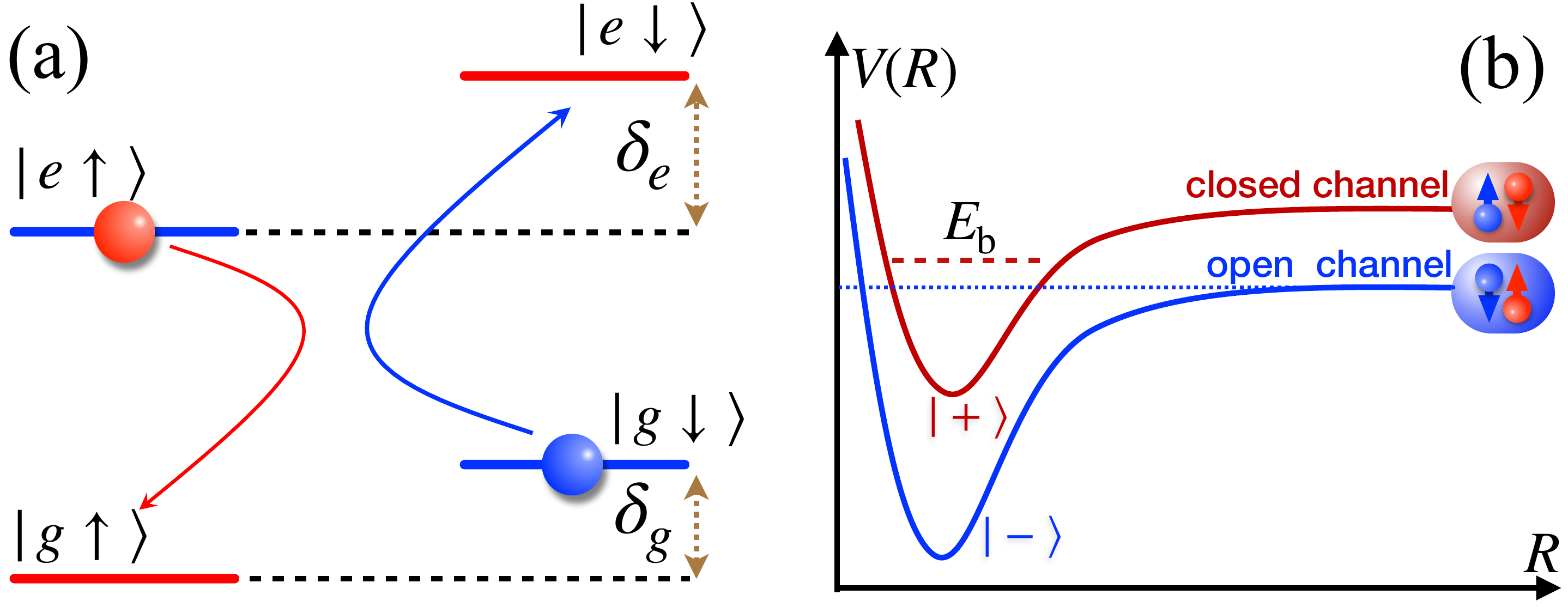}
\end{center}
\caption{ (a): Single atom energy level diagram of alkaline-earth(AE) atoms. Ground state $^{1}S_{0}$ and clock state $^{3}P_{0}$ are denoted by $|g\rangle$ and $|e\rangle$, respectively. Two out of N nuclear spin states are symbolized by $\uparrow$ and $\downarrow$. $\delta_{g}=Bg_{g}\mu_{B}$ and $\delta_{e}=Bg_{e}\mu_{B}$ are Zeeman energies of $|g\rangle$ and $|e\rangle$-atom, respectively. The energy difference $\delta=\delta_{e}-\delta_{g}$ between channels can be tuned by a magnetic field. The arrows show that two atoms initially occupying the open channel are scattered into the ``closed" channel. This process is also the spin-exchanging process.  (b): Interaction potentials between two AE atoms. At short range, the interaction potential is diagonal in the basis $|\pm\rangle$. Here $|\pm\rangle$ is the orbital singlet and orbital triplet, respectively. This is a major difference comparing to magnetic FR in alkaline-metal atom where interaction potentials at short distance are electron spin singlet and triplet. At long range, the interaction disappears and the closed and open channels are defined by eigenstates of free Hamiltonian.  When a bound state in the $|+\rangle$ channel is tuned near the scattering threshold, an OFR takes place.\label{fig1}}
\end{figure*}

The fermionic AE atoms have both the electronic orbital and the nuclear spin degree of freedoms. We denote the two-orbital states ${}^1S_0$ and ${}^3P_0$ as $|g\rangle$ and $|e\rangle$, respectively. For simplicity, we focus on two out of $N$ nuclear spin states, and symbolize them by $|\uparrow\rangle$ and $|\downarrow\rangle$, respectively. There are totally four states involved in the discussion and their energy levels are shown in Fig. \ref{fig1}(a). For single atom the nuclear spin degree of freedom and the orbital degree of freedom are decoupled because of $J=0$ in both states. When two atoms are far separated, let us consider two different single particle eigen-states, defined by
\begin{align}
&|\alpha\rangle=\frac{1}{\sqrt{2}}\left(|g\uparrow;e\downarrow\rangle-|e\downarrow;g\uparrow\rangle\right)\\
&|\beta\rangle=\frac{1}{\sqrt{2}}\left(|g\downarrow;e\uparrow\rangle-|e\uparrow;g\downarrow\rangle\right).
\end{align}
Here, for instance, the notation $|g\uparrow;e\downarrow\rangle$ denotes that the first atom is in the orbital-$g$ and nuclear spin-$\uparrow$ state, and the second atom is in the orbital-$e$ and nuclear spin-$\downarrow$ state. The wave function is anti-symmetrized because of the Fermi statistics. In the absence of the magnetic field, the energies of $|\alpha\rangle$ and $|\beta\rangle$ states are degenerate, and the degeneracy is broken by the magnetic field because of a slight difference in term of the Land\'e g-factor between different orbital states, arising from the higher order effect \cite{JunYePRA}. In term of the relative motion between two atoms, the non-interacting Hamiltonian when two atoms are far separated is given by 
\begin{equation}
\hat{H}_0=\left(-\frac{\hbar^2\nabla^2}{m}+\delta\right)|\alpha\rangle\langle\alpha|-\frac{\hbar^2\nabla^2}{m}|\beta\rangle\langle\beta|, \label{H0}
\end{equation}
where $\delta=\Delta m B\mu_B\delta g$, and $B$ is the magnetic field strength, $\Delta m$ is the difference of the nuclear spin quantum number along $\hat{z}$ direction between $|\uparrow\rangle $ and $|\downarrow\rangle$, $\mu_B$ is the Bohr's magneton and $\delta g$ is the difference in the Land\'e-g factor between two orbitals \cite{JunYePRA}. 

The general structure of the interaction part can be determined based on the nuclear spin rotational symmetry. Because of this symmetry, we should construct bases that are also invariant under the nuclear spin rotation, such that the interaction can be diagonal in these bases. These bases are defined as
\begin{equation}
|\pm\rangle=\frac{1}{2}(|ge\rangle\pm |eg\rangle)(|\uparrow\downarrow\rangle\mp |\downarrow\uparrow\rangle),
\end{equation}
where the nuclear spin part is either a singlet state or a triplet state that are spin rotational invariant. Here, since we only focus on the $s$-wave interaction where the spatial wave function is symmetric, the internal state wave function has to be anti-symmetric, therefore, the orbital part has to be a triplet (singlet) if the nuclear spin part is singlet (triplet). We define the projection operator $\mathcal{P}_{\pm}$ as $\mathcal{P}_{\pm}=|\pm\rangle\langle \pm|$, the interaction potential can be written as 
\begin{equation}
V({\bf r})=V_{+}({\bf r})\mathcal{P}_{+}+V_{-}({\bf r})\mathcal{P}_{-}. 
\end{equation}
The two interaction potential curves are schematically shown in Fig. \ref{fig1}(b). Generally speaking, $V_{\pm}({\bf r})$ are quite different because the orbital part of the wave functions are different for $|\pm\rangle$ channels. $H_0+V({\bf r})$ defines the total Hamiltonian for the two-body problem of AE atoms. 

Noticing that $|\pm\rangle$ can be related to $|\alpha\rangle$ and $|\beta\rangle$ via 
\begin{equation}
|\pm\rangle=\frac{1}{\sqrt{2}}(|\alpha\rangle\pm |\beta\rangle),
\end{equation}
the interaction term can be rewritten as
\begin{align}
V({\bf r})=\frac{V_{+}({\bf r})+V_{-}({\bf r})}{2}(|\alpha\rangle\langle\alpha|+|\beta\rangle\langle\beta|)\nonumber\\
+\frac{V_{-}({\bf r})-V_{+}({\bf r})}{2}(|\alpha\rangle\langle\beta|+|\beta\rangle\langle\alpha|). \label{Vr}
\end{align}
Here the interaction in the first term do not change the nuclear spin of atoms in $g$- and $e$-orbitals, and therefore it is considered as the spin-independent interaction between two orbitals. The second term exchanges the nuclear spin between atoms in $g$- and $e$-orbitals, and it is considered the spin-exchanging interaction, which is schematically shown by arrows in Fig.~\ref{fig1}(a). Below we will discuss how to control these two terms in two different sections, respectively. 

\section{Control of Spin-Independent Interaction \label{OFR}}

In the Box, we have discussed three essential ingredients that can lead to a FR. In fact, the Hamiltonian discussed above already shows that these conditions can also be satisfied. First, when two atoms are separated, $|\alpha\rangle$ and $|\beta\rangle$ are eigenstates of non-interacting part of the Hamiltonian, thus, they are regarded as the open and the closed channels, respectively. Secondly, the $\delta$-term in Eq. \ref{H0} means that the relative energy between two channels can be tuned by the magnetic field strength. Thirdly, the second term in Eq. \ref{Vr} means that the two channels are coupled at the short distance by the interaction potential, or more precisely, by the difference of two interaction potentials. This is schematically shown in Fig. \ref{fig1}(b).  

With these three conditions satisfied, one would expect that a FR can naturally exist between two AE atoms in different orbitals. However, there is one important obstacle. Since the two channels $|\alpha\rangle$ and $|\beta\rangle$ only involve the nuclear spin and the orbital degree of freedoms, the energy difference between these two channels depends on the nuclear magneton and the difference of the Land\'e-g factors between two orbitals, which are both very small \cite{JunYePRA}. Consequently, by increasing the magnetic field one Gauss, typically $\delta$ only increases by a few hundred Hz. This is much smaller comparing to the energy difference between the open and the closed channels of the alkaline-metal atoms, which is defined by the electronic spin degree of freedom and typically can increase by order of MHz for each Gauss. Therefore, for accessible magnetic field strength in cold atom laboratories, typically within a thousand Gauss, the tunable range of the energy difference between two channels is within $\sim 10^5$ Hz. That is to say, unless there exists a shallow bound state whose energy is within this range, otherwise the FR is not accessible.  

\begin{figure*}
\begin{center}
\includegraphics[width=0.4\textwidth]{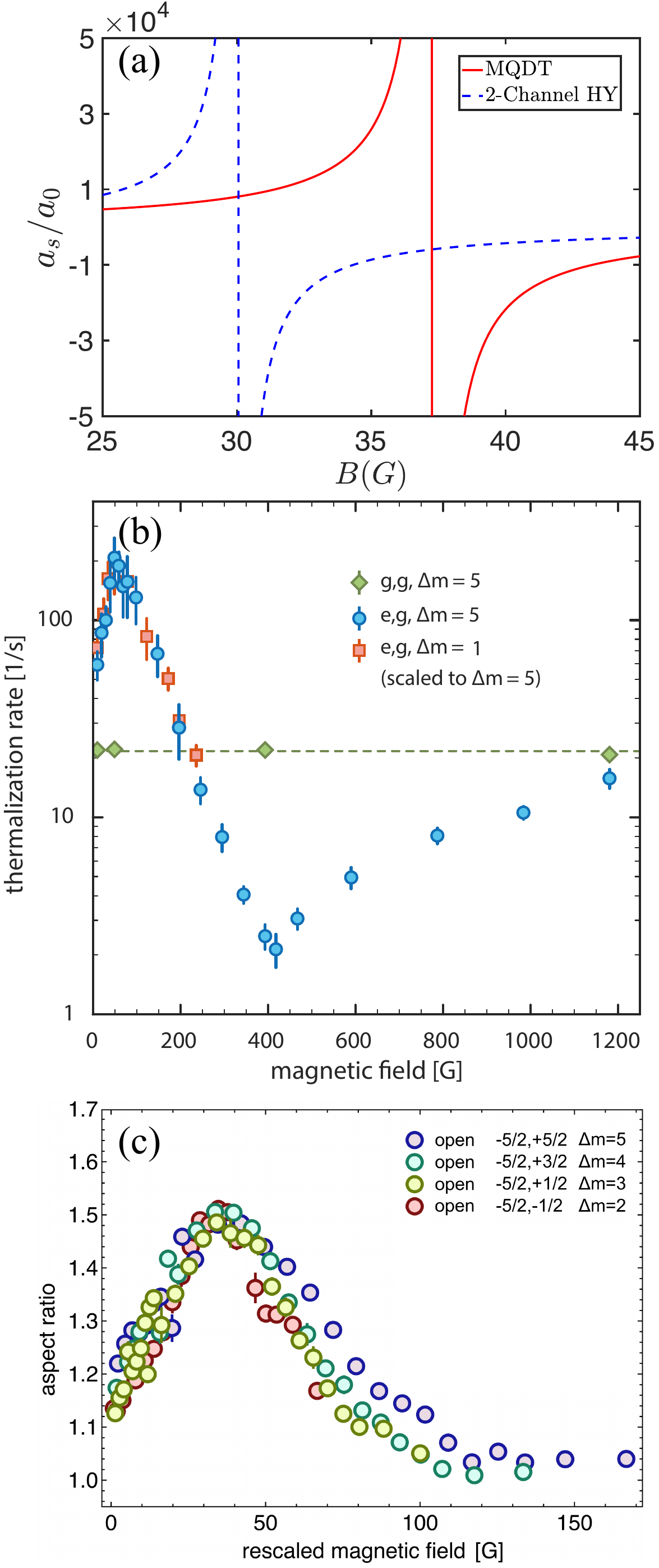}
\end{center}
\caption{Theoretical proposal and experimental demonstration of orbital Feshbach resonance(OFR) in $^{173}$Yb atoms. (a): The open channel scattering length as function of magnetic field. For $^{173}$Yb atoms, calculations with both the Huang-Yang(HY) pseudo-potential and the multi-channel quantum defect theory demonstrate that OFR appears around 40G of magnetic filed for $\Delta m=5$, which can be easily accessed in current experiment. (b): In the Munich experiment, the thermalization rate of $^{173}$Yb Fermi gas is measured. Circles and diamonds mark the thermalization rate of the $|e\rangle|g\rangle$ and $|g\rangle|g\rangle$ mixture with $\Delta m=5$, respectively. Square indicates the value of the $|e\rangle|g\rangle$ mixture with $\Delta m=1$ and the magnetic field axes has been scaled by $B\to B/5$. The sharp peak at $B\approx40$G indicates the existence of OFR. (b): In the Florence experiment, the aspect ratio of strongly interacting $^{173}$Yb Fermi gas is measured.  Aspect ratios as a function of scaled magnetic field for the atomic cloud with different $\Delta m$ are plotted together. The peak at $B\approx 40$G serves as another smoking gun evidence of the OFR. Figure (a), (b) and (c) are reprint from Ref.\cite{OFR}, \cite{OFR_exp1} and \cite{OFR_exp2}, respectively.\label{fig2}}
\end{figure*}

Fortunately, nature is very kind. It happens that $^{173}$Yb has a shallow bound state whose energy is about $\sim 4$ kHz. Hence, a FR is accessible in realistic magnetic field strength. Such a resonance is predicted first by theoretical calculation with pseudo-potential model \cite{OFR} and later has also been confirmed by the multi-channel quantum defect theory \cite{QDTOFR}. The predicted open channel scattering length is shown in Fig.~\ref{fig2}(a). To emphasize the important distinction between this resonance and the magnetic FR in alkaline-metal atoms, it is named as ``orbital Feshbach resonance" (OFR)\cite{OFR}, because here the channel is defined in term of the orbital degree of freedom while for FR in alkaline-metal atoms, channels are defined in term of the electronic spin degree of freedom. 

This theoretical prediction is soon confirmed by two experiments from the Munich group \cite{OFR_exp1} and the Florence group \cite{OFR_exp2}. In the Munich experiment \cite{OFR_exp1}, they prepare a mixture of $|g\uparrow\rangle$ and $|e\downarrow\rangle$, and measure the thermalization rate. A resonant interaction will lead to an unitary scattering cross section, and consequently, a very fast thermalization \cite{Jin2002}. Indeed, they find strong magnetic field dependence of the thermalization rate, with a peak at around $40$G consistent with the theoretical prediction, as shown in Fig.~\ref{fig2}(b). They also find a minimum at around $400$G, which confirms the existence of a zero-crossing of the scattering length due to the existence of the OFR. Moreover, they have checked that for single component $|g,\uparrow\rangle$ gas, there is no magnetic field dependence of the thermalization rate. 

In the Florence experiment \cite{OFR_exp2}, they confirm the existence of OFR by observing the emergent hydrodynamic behavior. Considering a degenerate Fermi gas confined in an anisotropic trap, if the gas is non-interacting or weakly interacting, the expansion dynamics is ballistic when the trap is turned off, and the atomic cloud will eventually become isotropic, with the aspect ratio approaching unity. However, when the gas is strongly interacting, the expansion dynamics will become hydrodynamic. The pressure is larger in the short direction so that the expansion will be faster. Therefore, the shorter axes before expansion will become longer axes after expansion. The inverted aspect ratio is taken as a hallmark of hydrodynamical behavior, which indicates the system is a strongly interacting one \cite{ohara2002}. Indeed, it has been observed that the aspect ratio of the gas after sufficiently long time expansion depends on the magnetic field strength, with the largest inverted aspect ratio also peaked at around $40$G, as shown in Fig.~\ref{fig2}(c).  

These two experiments successfully confirmed OFR in ${}^{173}$Yb atom. An extra feature worth emphasizing here is the nuclear spin dependence of the resonance. The nuclear spin of ${}^{173}$Yb is $5/2$ and there are totally six nuclear spin states. In the discussion above, we take any two out of six states and label them as $|\uparrow\rangle$ and $|\downarrow\rangle$, and $\Delta m$ can take any integer from one to five. As we have discussed in Sec. \ref{Introduction}, the interaction potential $V_{\pm}({\bf r})$ should be independent of the choice of nuclear spin, that is to say, they should be the same for any combinations. Hence, the only nuclear spin dependence comes from $\Delta m$ in the expression of $\delta$. Therefore, if we plot the observables in term of scaled magnetic field strength $B/\Delta m$, the measurement with different nuclear spin combinations should collapse into the same curve. Such a scaling behavior has indeed be observed in both thermalization experiment by the Munich group and aspect ratio measurement by the Florence group \cite{OFR_exp1,OFR_exp2}.

Finally, it is very important to note that both experiments find that the lifetime of a degenerate ${}^{173}$Yb Fermi gas nearby the OFR can be as long as a few hundred milli-second, which is sufficiently long to reach many-body equilibrium \cite{OFR_exp1,OFR_exp2}. A many-body system nearby a FR is a strongly interacting one with many intriguing properties, such as BEC-BCS crossover, universal thermodynamics, high transition temperature, and pseudo-gap physics \cite{Fermigas_review,Strinati}. Such systems have been extensively studied using magnetic FR in alkali-metal atoms. A natural equation is, therefore, to ask what is the difference in many-body physics between OFR and magnetic FR in alkaline-metal atoms. The answer to this question again lies in the fact that the channels here are defined in term of the ``orbitals" instead of the electronic spin. We will discuss it in detail in the Sec.\ref{Outlook}.

\section{Control of Spin-Exchanging Interaction \label{Spin}}

The coupling between $|\alpha\rangle$ and $|\beta\rangle$ channels described by the second term in Eq. \ref{Vr} gives rise to a nuclear spin exchanging processes between two orbitals. This spin-exchanging phenomenon has been observed by the Munich group \cite{spin-exch1} and the Florence group \cite{spin-exch2}. For this process to be prominent, these two channels should be degenerate in term of single-particle energy. Hence, in this section, we will mainly focus on the zero magnetic field case with $\delta=0$, and this is in contrast to Sec. \ref{OFR} where the magnetic field strength is always finite. Note that the spin-exchanging interaction depends on the difference between $V_{+}({\bf r})$ and $V_{-}({\bf r})$, therefore, the key point is that if at a certain circumstance, one of $V_{\pm}({\bf r})$ gets amplified by resonant interaction but the other does not, the spin-exchanging interaction will be enhanced \cite{CIR1}.  

\begin{figure*}
\begin{center}
\includegraphics[width=0.4\textwidth]{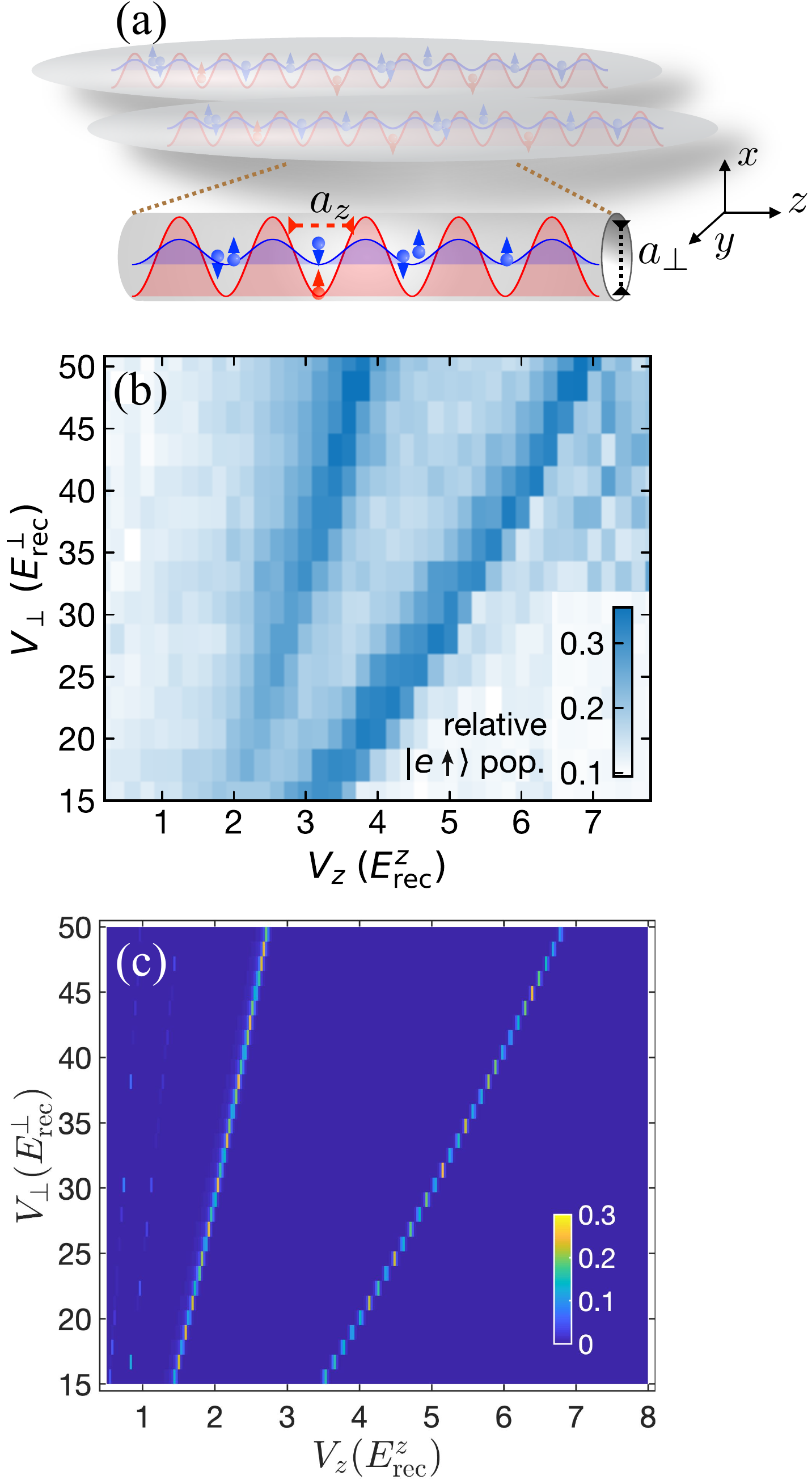}
\end{center}
\caption{Theoretical proposal and experimental demonstration of controlling spin-exchanging interaction. (a): Illustration of experimental set up for realization of the confinement-induced resonance(CIR) in alkaline-earth atoms to tune the spin-exchanging interaction. The $^{1}S_{0}$ state (blue ball) and $^{3}P_{0}$ state (red ball) of alkaline-earth atoms are simultaneously trapped in 2D optical lattice in the $xy$ plane by a laser at the magic wavelength laser.  An orbit-dependent laser is applied along the $z$ direction, which creates a deep lattice for localizing $|e\rangle$ atoms and a shallow lattice for $|g\rangle$ atoms to keep it mobile. (b): Experimental measurement of spin-exchanging scattering amplitude as function of the transverse and the longitudinal lattice depths. The scattering amplitude is determined by measuring the population of $|e,\uparrow\rangle$ state. The observed two resonances are originated from the CIRs. (c): Theoretical results of the spin-exchanging scattering amplitude of the same system and the same parameter regime as the experiments. This result quantitatively agree with the experimental observation. Figure (b) and (c) are reprinted from Ref.\cite{CIR_exp} and \cite{CIR4}, respectively.\label{fig3}}
\end{figure*}

This idea can be implemented by utilizing CIR. For $\delta=0$, the non-interacting part is also diagonal in $|\pm\rangle$ bases, and since the interaction term is also diagonal in $|\pm\rangle$ bases, $|+\rangle$ and $|-\rangle$ become two decoupled channels and can be treated independently. Generally, the two states ${}^1S_0$ and ${}^3P_0$ have difference ac polarization and experience different potential in a laser field. However, there exist magnetic wave lengths at which these two states incidentally have the same ac polarization \cite{magic1,magic2}. One can create a deep two-dimensional optical lattice with a magic wavelength laser, such that atoms can move freely along a one-dimensional tube with the same transverse confinement length $a_\perp$ for both two orbitals, as shown in Fig.~\ref{fig3}(a). According to the discussion in the Box, the $|\pm\rangle$ channel reaches CIR when $a_{\perp}=\mathcal{C}a_{\pm}$ \cite{oshanii98}, where $a_{\pm}$ are the scattering length for $V_{\pm}({\bf r})$ potential, respectively. Because $a_{\pm}$ are different, only one of these two channels can reach CIR either at $a_{\perp}=\mathcal{C}a_{+}$ or at $a_{\perp}=\mathcal{C}a_{-}$, and therefore, the spin-exchanging interaction will get amplified. This is the basic idea of how to control spin-exchanging interaction with CIR \cite{CIR1}. 

In practices, when the laser wavelength is fixed at the magic wavelength, $a_\perp$ can only be tuned by the depth of the two-dimensional optical lattice $V_\perp$, and the tunable range is limited. One of the major goals for enhancing the spin-exchanging interaction is to realize the Kondo effect, which is caused by localized magnetic impurities embedded in a mobile Fermi sea \cite{kondobook}. Hence, we further consider the situation that atoms in the ${}^3P_0$ state are localized by a deep lattice to a zero-dimension one and atoms in ${}^1S_0$ state experience a shallow lattice along the one-dimensional tube and remain mobile. This situation is realized by imposing another optical lattice along the tube direction and this laser is \textit{not} at the magic wavelength, such that the lattice potential $V_z$ for ${}^3P_0$ state is sufficiently deep but the lattice potential for ${}^1S_0$ state is shallow. In the theoretical model, it is a good approximation to consider atoms in ${}^3P_0$ state as localized by a harmonic trap along the tube, with the harmonic length denoted by $a_z$ and controlled by $V_z$. Therefore, $a_\perp$ and $a_z$, or equivalent to say, $V_\perp$ and $V_z$, are two tuning parameters, with which the tunable range is broad enough to reach a CIR for practical parameters \cite{CIR2}. This $1+0$-dimensional CIR problem with these two parameters has now been studied both theoretically and experimentally. As illustrated in Fig.~\ref{fig3}(b), experimentally this situation is realized by the Munich group and a resonant enhanced spin-exchanging scattering amplitude has been observed \cite{CIR_exp}. Theoretically, this problem can be treated with different degrees of approximations \cite{CIR1,CIR2,CIR3,CIR4,qingjiCIR}, and quantitative results can be systematically improved \cite{CIR4}. The most accurate results are shown in Fig.~\ref{fig3}(c) that are compared with the experimental results quite well.   

\section{Outlook \label{Outlook}} 
Above we have discussed proposals and experiments of how we can control both the spin-independent and spin-exchanging interaction in ultracold AE atoms. Here we will discuss several aspects of novel physics that can emerge from these interaction controls. 

\textbf{1. The strongly interacting many-body systems nearby the OFR display novel features.} For a typical magnetic FR in the alkaline-metal atom, the energy separation between open and closed channels is of the order of $10^{8}-10^{9}$ Hz, which is a few orders of magnitude larger than the typical Fermi energy of a degenerate fermionic atomic gas. Therefore, such a system can be either described by a single open channel only, or by two coupled channels but only including the bound state in the closed channel. The closed channel has never been populated by scattering states, and that is the reason it is called ``closed". However, the situation is completely different in the OFR case. As we have explained in Sec. \ref{OFR}, because the two channels are defined in term of the orbital degree of freedom, the energy separation between them is small, and it is comparable or even smaller than the Fermi energy for a typical density of degenerate Fermi gas. This forces us to take into account the scattering state in the closed channel, that is to say, the so-called closed channel is not really ``closed". The multi-orbital nature brings the physics into a new parameter regime that has not been explored before \cite{OFR,junjun,lianyi,iskin,weiyi,yicai,iskin2,tianshu,xiaofan,xiaofan2,yoji1,yoji2,yoji3,haiyang,klimin,zoupeng,laird}. Here we simply give couple examples. First of all, the strong influence of the closed channel leads to a strong momentum dependence of the scattering amplitude, which can lead to an even higher transition temperature for forming Fermi superfluid, comparing to the single-channel wide resonance in alkaline-metal atom \cite{junjun}. In the superfluid phase, since both two channels are occupied by the scattering states, it requires two order pairing order parameters to describes such a Fermi superfluid, which adds a new twist to the BEC-BCS crossover physics \cite{OFR,junjun,iskin,iskin2,yoji1,yoji2,yoji3,klimin,laird}. This is reminiscent of multi-band superconductor in solid-state materials. A novel collective mode known as the Leggett mode is predicted to exist in such situation, which describes the relative phase fluctuation between two order parameters \cite{lianyi,yicai}. Realizing such a multi-order-parameter Fermi superfluid provides a platform to observe this long-sought collective mode. Recently, the Florence experimental group has directly observed and coherently manipulate the orbital Feshbach molecule, which is the first and important step towards the multi-order-parameter Fermi superfluid \cite{OFR-molecule}.

\textbf{2. An important application of OFR is to realize topological superfluid.} Topological superfluid can be realized by introducing spin-orbit coupling into Fermi superfluid \cite{topological_SF}. Since currently, the lowest temperature that can be reached in degenerate fermionic atom gas is about $0.1$ times of the Fermi temperature, one has to rely on the FR to reach Fermi superfluid. On the other hand, in cold atomic gas spin-orbit coupling can be simulated by coupling two internal spin states with a laser and the coupling is accompanied by momentum transfer. Therefore, to realize topological superfluid in cold atomic gases, a system needs to be equipped with both spin-orbit coupling and FR simultaneously. 

In alkaline-metal atoms, there are many well-behaved FRs. The spin-orbit coupling can also be simulated by coupling two hyperfine spin states with a Raman beam. However, such a Raman transition leads to heating due to the spontaneous emission of the intermediate state. This heating so far prevents reaching Fermi superfluid in the presence of spin-orbit coupling. In lanthanide atoms, the spin-orbit coupling can also be realized by Raman coupling and heating can be suppressed, as shown in ${}^{164}$Dy atoms \cite{xiaolingDy}. However, the FRs observed in lanthanide so far are all too narrow to realize Fermi superfluid. Moreover, lanthanide atoms usually have a large spin and therefore the dipolar loss is also an obstacle \cite{Dysoc}. 

In AE atoms, there are two ways to realize spin-orbit coupling, either by coupling two nuclear spin state in ${}^1S_0$ manifold  via Raman coupling \cite{ground_soc1,ground_soc2,ground_soc3,ground_soc4} or by directly coupling ${}^1S_0$ and ${}^3P_0$ states by a laser \cite{socclock1,socclock2}. Here we focus on the later because in this setting, there is no inter-mediate state involved and heating due to spontaneous emission can be avoided, which has been shown in both ${}^{88}$Sr and ${}^{173}$Yb atoms \cite{socclock1,socclock2}. Nevertheless, if there is no FR between these two states, topological superfluid is still not possible. Thus, discovering OFR in ${}^{173}$Yb fills the gap and is therefore very crucial. In view of these discussions, ${}^{173}$Yb can be the best candidate for realizing topological superfluid in ultracold atomic gases.  

\textbf{3. AE atomic gas, in particular, cold ${}^{173}$Yb atom gas, provides a unique platform for studying impurity physics.} Impurity physics is an important subject in quantum many-body physics. For spin-independent interaction, a moving impurity can be dressed up the background particle and forms a quasi-particle called ``polaron" \cite{chevy,cui_polaron}. In the Fermi liquid phase nearby FR, polaron is an important elementary excitation related to issues like the stability of ferromagnetism and the phase diagram of imbalanced Fermi gas \cite{chevy, cui_polaron,polaron-review}. For spin-exchanging interaction, as already discussed in Sec. \ref{Spin}, a static impurity in a Fermi sea can lead to the famous Kondo effect\cite{kondobook}. In previous studies in condensed matter materials and ultracold alkaline-metal atomic gases, both two types of impurity physics have been studied separately. It is interesting to note that both two types of impurity physics can exist in the system of two orbital mixture of cold ${}^{173}$Yb gases \cite{jinge,qiran,jinge2,tianshu2,polaron-exp}. By tuning the magnetic field strength from zero to finite, and by tuning the confinement potential, the interaction can be tuned from the spin-exchanging interaction dominated regime to the spin-independent interaction dominated regime; and by tuning the external lattice potential, one can tune the impurity from localized to mobile. Thus, one may have a chance to establish a coherent picture for the crossover from polaron physics to Kondo physics.  

\textbf{4. Realizing Kondo effect with cold atomic gas can enrich our understanding of Kondo physics itself.} There is an increasing interests of simulating Kondo physics with ultracold atoms \cite{kondo1,kondo2,kondo3,kondo4,kondo5,kondo6,kondo7,kondo8,kondo9,kondo10,kondo11,kondo12,kondo13,kondo14,kondo15}. First of all, for the Kondo effect observed in condensed matter materials, the spin-exchanging coupling is usually much weaker than the Fermi energy and therefore the Kondo temperature is also much lower than the Fermi temperature. Since in ultracold atomic gas, we cannot reach a temperature much lower than the Fermi energy, we have to increase the spin-exchanging interaction to enhance the Kondo temperature to an attainable regime. This brings the Kondo physics to a strong coupling regime where the Kondo coupling can be comparable to the Fermi energy. This is somewhat reminiscent of Fermi superfluid at unitarity where pairing energy is comparable to the Fermi energy. Secondly, as we mentioned in the beginning, the fermionic AE atoms usually have a large nuclear spin and the interaction displays $SU(N)$ symmetry. This can naturally realize the $SU(N)$ Kondo model that cannot be easily realized in solid-state materials before. Thirdly, because in ultracold atom system we can control the atom number precisely from a few to many, and previous experiments have studied how the pairing correlation is established from few-body to many-body system by adding one particle at a time \cite{few2many}, and with similar scheme one can also study how the Kondo screening is built up from few-to-many perspective. Finally, it is well known that ultracold atom systems are good for study non-equilibrium dynamics such as quench dynamics because the parameters can be changed in a time scale much faster than many-body equilibrium time. One can take this advantage to study non-equilibrium dynamics related to the Kondo physics \cite{kondo13,kondo14}, for instance, by suddenly quench the spin-exchanging interaction from ferromagnetic to the anti-ferromagnetic regime, one can study the dynamical formation of the Kondo screening.    

\section{Conclusion.}  We summarize this perspective as Table \ref{table}. We review recent developments of controlling both spin-independent and spin-exchanging interaction in ultracold AE atoms. We highlight that these tools of controlling interactions, combined with the two-orbital physics and $SU(N)$ physics of AE atoms given by their natural atomic structure, can open up new front of quantum simulation with ultracold atoms, including but not limited to, studying novel type of strongly interacting Fermi gas, realizing topological superfluid and investigating far-from-equilibrium and $SU(N)$ aspects of the Kondo physics. 

\textit{Acknowledgement.} This work is supported MOST (Grant No. 2018YFA0307601 (RZ), 2018YFA0306502 (PZ),  2016YFA0301600 (HZ)) and NSFC (Grant No.11804268 (RZ), 11434011(PZ), 11674393(PZ), 11734010 (HZ)). Peng Zhang is also supported by the Research Funds of Renmin University of China (Grant No. 16XNLQ03).

\begin{widetext}

\begin{table}[h] 
\centering
\begin{tabular}{|>{\centering\arraybackslash}m{2.8cm}|>{\centering\arraybackslash}m{3.0cm}|>{\centering\arraybackslash}m{3.5cm}|>{\centering\arraybackslash}m{3cm}|>{\centering\arraybackslash}m{5cm}|} 
\hline
Interaction Types & Definition & Control Tool &Magnetice Field Regime & Applications\\ 
\hline 
Spin-Independent Interaction& Diagonal Term in Eq.(\ref{Vr}) & Orbital Feshbach Resonance(OFR) & Finite Field & Strongly Interacting Fermi Gases;\  Topological Superfluid, \dots \\
\hline 
Spin-Exchanging Interaction & Off-Diagonal Term in Eq.(\ref{Vr}) & Confinement-induced Resonance(CIR) &Zero Field & $SU(N)$ and Non-Equilibrium Kondo Physics, \dots \\
\hline
\end{tabular}
\caption{Summary of controlling two types of interaction in AE atoms between two orbitals. }
\label{table}
\end{table}

\end{widetext}

\begin{widetext}
\centering
\fbox
{\parbox{0.95\textwidth}
{
{\bf Box1:  Feshbach Resonance}

Feshbach resonance (FR) refers to a type of scattering resonance between two atoms. Generally speaking, there are three ingredients in order to realize a FR. 
 
 \begin{enumerate}
  \item There have to be at least two interaction channels defined by different eigenstates when two atoms are far separated from each other. 
  \item The energy difference between these two channels can be tuned by an external control parameter. 
  \item At a short distance, these two channels are coupled. 
\end{enumerate}

Usually, when two atoms are far separated, they both stay at their single-particle eigenstate. The channel with lower threshold energy is called the open channel, and the one with higher threshold energy is called the closed channel. When the relative energy between the open and the closed channel is tuned to match a bound state energy in the closed channel, this bound state appears at the threshold of the open channel, which leads to a scattering resonance due to their coupling at a short distance. The interaction between two atoms can, therefore, be tuned by controlling the relative energy between these two channels, and the interaction becomes the strongest at the resonance. 

For the magnetic FR of alkaline-metal atoms, the three conditions are satisfied as follows: (i) When two atoms are far separated, each atom stays in the eigenstate of internal degrees of freedoms, determined by the hyperfine coupling and the Zeeman energy. (ii) The energy difference between these two channels can be tuned by the Zeeman energy controlled by the external magnetic field. (iii) At short-distance, the interaction potential is largely determined by the total electronic spin of two atoms, which does not commute with quantum number of single-particle eigenstate, and therefore two channels are mixed. 

The confinement-induced-resonance (CIR) can also be understood in term of the same mechanism as FR \cite{oshanii98,oshanii03}. For instance, considering two atoms in a one-dimensional tube, the three conditions can also be satisfied as follows: (i) When two atoms are far separated along the longitudinal direction, atoms can be in different eigenmodes in the transverse direction. (ii) The energy difference between these two channels can be controlled by the strength of the transverse confinement. (iii) At short-distance, the interaction potential can scatter atoms from one transverse mode to another one, and therefore couples two channels. That is to say, the transverse confinement plays the role as the internal degree of freedoms in the magnetic FR. Since the energy separation between two transverse modes is $2\hbar^2/(ma^2_\perp)$, where $a_\perp$ is the transverse harmonic length, and the bound state energy can be estimated as $\hbar^2/(ma^2_\text{s})$, where $a_\text{s}$ is the $s$-wave scattering length, CIR occurs when these two energies match each other that gives $a_\text{s}=a_\perp/\sqrt{2}$. A rigorous calculation taking into account contributions from all high transverse modes gives precise results that CIR occurs at $a_\text{s}=a_\perp/\mathcal{C}$, with $\mathcal{C}=1.4603\cdots$. 
}
}

\end{widetext}

%***********************************
%***********************************

%***********************************
%***********************************
%***********************************
%***********************************


\begin{thebibliography}{99}
\bibitem{Ybnote}
Although Ytterbium (Yb) atom is not AE atom in the periodic table, it has the same out-shell electronic structure as the AE atoms, therefore our discussion will include Yb atoms.
\bibitem{socfootnote}
To classify the terminology of spin-orbit coupling, here the internal orbital degree of freedom is used to simulate ``spin" and ``orbit" means the spatial motion of atoms.
\end{thebibliography}
\end{document}